\documentclass[]{ceurart}

\usepackage{amsmath,amssymb}
\usepackage{booktabs}
\usepackage{graphicx}
\usepackage{tikz}
\usepackage{pgfplots}
\pgfplotsset{compat=1.18}
\usetikzlibrary{shapes.geometric, arrows.meta, positioning, fit, calc, patterns}
\usepackage{multirow}
\usepackage[table]{xcolor}
\usepackage{colortbl}
\usepackage{enumitem}
\usepackage{subcaption}

\begin{document}

\copyrightyear{2026}
\copyrightclause{Copyright for this paper by its authors.
  Use permitted under Creative Commons License Attribution 4.0
  International (CC BY 4.0).}
\conference{ECOM'26: SIGIR Workshop on eCommerce, Jul 24, 2026, Melbourne, Australia}

\title{Unified Multi-Task Relevance Modeling for E-Commerce: Comparing Task Routing Architectures Across LLMs and Cross-Encoders}

\author[1]{Md Omar Faruk Rokon}[%
  email=mdomarfaruk.rokon@walmart.com,
]
\cormark[1]
\author[1]{Jhalak Nilesh Acharya}[%
  email=jhalak.acharya@walmart.com,
]
\cormark[1]
\author[1]{Shasvat Desai}[%
  email=shasvat.desai@walmart.com,
]
\cormark[1]
\author[1]{Hong Yao}[%
  email=hong.yao0@walmart.com,
]
\author[1]{Kuang-chih Lee}[%
  email=kuang-chih.lee@walmart.com,
]

\address[1]{Walmart Global Tech, Sunnyvale, CA, USA}

\cortext[1]{Corresponding author.}

\begin{abstract}
How can we build a single relevance model that handles six different entity-pair relationship types in e-commerce---from query--product matching to product-type similarity---when each task has different data volumes, different semantic requirements, and potentially conflicting learning signals? This question is important because current industry practice relies on separate models for each task, preventing knowledge transfer and producing inconsistent relevance signals. Our work is driven by the following insight: encoder-based and decoder-only models encode task identity through different mechanisms, so the choice of task routing architecture---how task identity is communicated to the shared model---affects these two families in asymmetric ways. As our key novelty, we combine three ideas: (a)~a unified multi-task framework that jointly trains on six entity-pair tasks under a shared three-point relevance scale, (b)~a systematic comparison of three task routing architectures (text-prefix routing, multi-head classification, and multi-head with private transformer layers) across both LoRA-adapted LLMs and fully fine-tuned cross-encoders, and (c)~a majority-vote ensemble that exploits the diversity induced by private-layer routing. First, we show that the MHP Ensemble (multi-head with private layers) achieves 89.96\% accuracy on 453K test examples---the highest across all configurations (+1.92\% over the text-prefix ensemble). Second, we show that removing text prefixes without private layers causes severe degradation for decoder-only LLMs (Gemma-2-2b: 87.14\% $\to$ 55.59\%) while cross-encoders remain robust ($-$1.02\%), suggesting an encoder--decoder asymmetry in task identity encoding. Third, we show that multi-task training yields up to +14\% improvement on low-resource tasks over single-task baselines. Overall, our work provides a practical blueprint for building unified multi-task relevance models in e-commerce, with actionable guidance on task routing architecture selection for different model families.
\end{abstract}


\maketitle

\section{Introduction}
\label{sec:intro}

Can a single model unify six distinct entity-pair relevance tasks in e-commerce---from query--product matching to product-type similarity---despite differences in data volume, semantic granularity, and potential conflicts in learning signal? This question is important because e-commerce platforms must evaluate relevance across query--product matching for search, query--query similarity for suggestions, product--product relatedness for recommendations, and query--product-type mappings for intent understanding~\cite{aiello2016relevance, su2018user, wang2023clickconv}. Current industry practice trains separate models for each task, which prevents knowledge transfer and produces inconsistent relevance signals across downstream systems.

Consider a concrete example. When a user searches for ``hiking boots,'' the platform must determine: (1)~whether the ad item ``Merrell Moab 3 Mid'' is relevant to the query, (2)~whether the query ``outdoor footwear'' is similar to ``hiking boots,'' (3)~whether ``hiking boots'' maps to the product type \emph{Hiking Boots}, and so on for six relationship types. Training a separate model for each task is wasteful---the query ``hiking boots'' carries the same semantic meaning regardless of which relationship is being evaluated. A single model that shares this understanding across tasks should transfer knowledge from data-rich tasks (query--ad: 400K examples) to data-scarce tasks (product-type--product-type: only 14K examples).

Our work is driven by the following insight: encoder-based and decoder-only models encode task identity through different mechanisms. Cross-encoders pool from a dedicated [CLS] token that summarizes the full input, making it relatively agnostic to how task identity is communicated. Decoder-only LLMs pool from the last token, whose representation is shaped by the causal attention mask and heavily influenced by the input prefix. When the task-identifying prefix is removed, these two families respond asymmetrically---cross-encoders are robust, but decoder-only LLMs collapse.

\begin{figure*}[t]
\centering
\small
\resizebox{\textwidth}{!}{%
\begin{tikzpicture}[
    phase/.style={rectangle, draw, rounded corners=3pt, minimum height=0.7cm, align=center, font=\small\bfseries, minimum width=2.6cm},
    comp/.style={rectangle, draw, rounded corners=2pt, minimum height=0.55cm, align=center, font=\scriptsize, minimum width=2.4cm},
    arr/.style={-{Stealth[length=2mm]}, thick},
    garr/.style={-{Stealth[length=1.5mm]}, semithick, gray!60},
]

\node[phase, fill=yellow!15] (data) at (0,0) {Dataset\\Construction};
\node[comp, fill=yellow!6] (seed) at (0,-1.2) {T1: Human-labeled\\800K Q--Ad pairs};
\node[comp, fill=yellow!6] (derived) at (0,-2.4) {T2--T6: Derived\\co-relevance + taxonomy};
\node[comp, fill=yellow!6] (final) at (0,-3.6) {2.27M examples\\6 tasks, 3 classes};
\draw[garr] (data) -- (seed);
\draw[garr] (seed) -- (derived);
\draw[garr] (derived) -- (final);

\node[phase, fill=orange!15] (train) at (5,0) {Multi-Task\\Training};
\node[comp, fill=orange!6] (lora) at (5,-1.2) {LoRA LLMs\\Gemma-2b, Llama-8b};
\node[comp, fill=orange!6] (xe) at (5,-2.4) {Cross-Encoders\\ModernBERT, ELECTRA};
\node[comp, fill=red!6] (routing) at (5,-3.6) {$\times$\,3 routing:\\SH, MH, MHP};
\draw[garr] (train) -- (lora);
\draw[garr] (lora) -- (xe);
\draw[garr] (xe) -- (routing);

\node[phase, fill=blue!12] (eval) at (10,0) {Evaluation\\453K test};
\node[comp, fill=blue!6] (metrics) at (10,-1.2) {Per-task accuracy\\Per-class P/R/F1};
\node[comp, fill=blue!6] (ablation) at (10,-2.4) {Ablation studies\\routing, scale, weights};
\draw[garr] (eval) -- (metrics);
\draw[garr] (metrics) -- (ablation);

\node[phase, fill=green!15] (ens) at (15,0) {Majority-Vote\\Ensemble};
\node[comp, fill=green!6] (v2ens) at (15,-1.2) {MHP Ensemble\\89.96\% accuracy};
\node[comp, fill=green!6] (deploy) at (15,-2.4) {Production\\deployment};
\draw[garr] (ens) -- (v2ens);
\draw[garr] (v2ens) -- (deploy);

\draw[arr] (data) -- (train);
\draw[arr] (train) -- (eval);
\draw[arr] (eval) -- (ens);

\draw[arr, gray!60] (final.east) -- ++(0.4,0) |- ([yshift=-2pt]train.south west);

\end{tikzpicture}%
}
\caption{End-to-end system overview: dataset construction from human-labeled seed data, multi-task training of LoRA LLMs and cross-encoders across three routing architectures, evaluation on 453K test examples, and majority-vote ensemble achieving 89.96\% accuracy.}
\label{fig:system_overview}
\end{figure*}
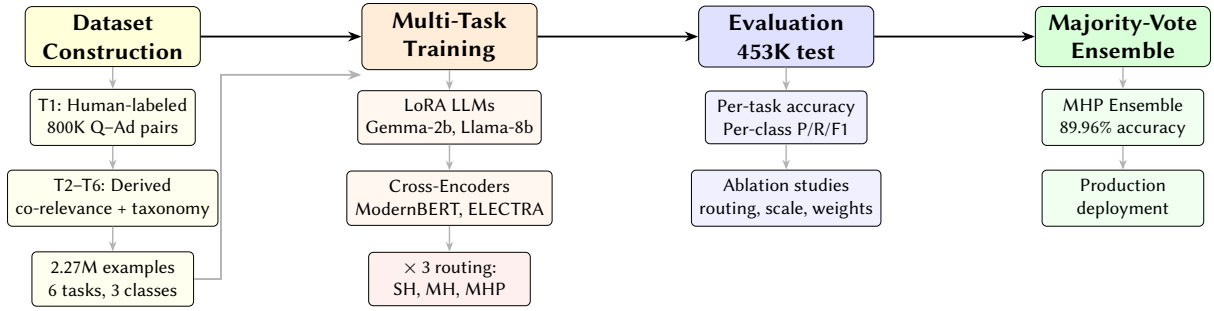

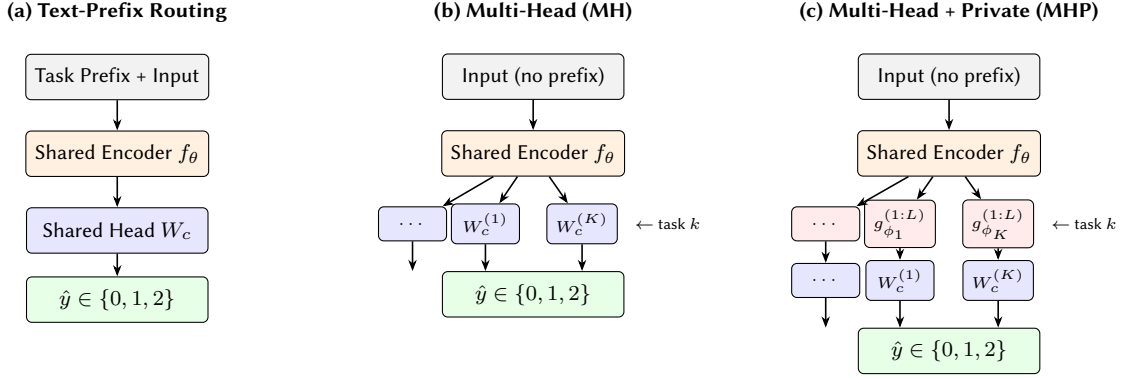
\begin{figure*}[t]
\centering
\small
\begin{tikzpicture}[
    node distance=0.5cm and 0.4cm,
    sbox/.style={rectangle, draw, rounded corners=2pt, minimum width=2.4cm, minimum height=0.6cm, align=center, font=\scriptsize},
    hbox/.style={rectangle, draw, rounded corners=2pt, minimum width=0.9cm, minimum height=0.45cm, align=center, font=\tiny},
    pbox/.style={rectangle, draw, rounded corners=2pt, minimum width=0.9cm, minimum height=0.45cm, align=center, font=\tiny, fill=red!8},
    arr/.style={-{Stealth[length=1.5mm]}, semithick},
    lbl/.style={font=\scriptsize\bfseries},
]

\node[lbl] (la) at (0, 0) {(a) Text-Prefix Routing};
\node[sbox, below=0.3cm of la, fill=gray!10] (pfx) {Task Prefix + Input};
\node[sbox, below=0.4cm of pfx, fill=orange!12] (enc_a) {Shared Encoder $f_\theta$};
\node[sbox, below=0.4cm of enc_a, fill=blue!10] (head_a) {Shared Head $W_c$};
\node[sbox, below=0.3cm of head_a, fill=green!10] (out_a) {$\hat{y} \in \{0,1,2\}$};
\draw[arr] (pfx) -- (enc_a);
\draw[arr] (enc_a) -- (head_a);
\draw[arr] (head_a) -- (out_a);

\node[lbl] (lb) at (5.5, 0) {(b) Multi-Head (MH)};
\node[sbox, below=0.3cm of lb, fill=gray!10] (inp_b) {Input (no prefix)};
\node[sbox, below=0.4cm of inp_b, fill=orange!12] (enc_b) {Shared Encoder $f_\theta$};
\node[below=0.5cm of enc_b] (mid_b) {};
\node[hbox, fill=blue!10, left=0.05cm of mid_b] (h1) {$W_c^{(1)}$};
\node[hbox, fill=blue!10, left=0.05cm of h1] (h0) {$\cdots$};
\node[hbox, fill=blue!10, right=0.05cm of mid_b] (h2) {$W_c^{(K)}$};
\node[sbox, below=0.5cm of mid_b, fill=green!10] (out_b) {$\hat{y} \in \{0,1,2\}$};
\draw[arr] (inp_b) -- (enc_b);
\draw[arr] (enc_b) -- (h0);
\draw[arr] (enc_b) -- (h1);
\draw[arr] (enc_b) -- (h2);
\draw[arr] (h0) -- (h0 |- out_b.north);
\draw[arr] (h1) -- (h1 |- out_b.north);
\draw[arr] (h2) -- (h2 |- out_b.north);
\node[font=\tiny, right=0.1cm of h2] {$\leftarrow$ task $k$};

\node[lbl] (lc) at (11, 0) {(c) Multi-Head + Private (MHP)};
\node[sbox, below=0.3cm of lc, fill=gray!10] (inp_c) {Input (no prefix)};
\node[sbox, below=0.4cm of inp_c, fill=orange!12] (enc_c) {Shared Encoder $f_\theta$};
\node[below=0.5cm of enc_c] (mid_c) {};
\node[pbox, left=0.05cm of mid_c] (p1) {$g_{\phi_1}^{(1:L)}$};
\node[pbox, left=0.05cm of p1] (p0) {$\cdots$};
\node[pbox, right=0.05cm of mid_c] (p2) {$g_{\phi_K}^{(1:L)}$};
\node[below=0.5cm of mid_c] (mid_c2) {};
\node[hbox, fill=blue!10] at (p0 |- mid_c2) (hc0) {$\cdots$};
\node[hbox, fill=blue!10] at (p1 |- mid_c2) (hc1) {$W_c^{(1)}$};
\node[hbox, fill=blue!10] at (p2 |- mid_c2) (hc2) {$W_c^{(K)}$};
\node[sbox, below=0.5cm of mid_c2, fill=green!10] (out_c) {$\hat{y} \in \{0,1,2\}$};
\draw[arr] (inp_c) -- (enc_c);
\draw[arr] (enc_c) -- (p0);
\draw[arr] (enc_c) -- (p1);
\draw[arr] (enc_c) -- (p2);
\draw[arr] (p0) -- (hc0);
\draw[arr] (p1) -- (hc1);
\draw[arr] (p2) -- (hc2);
\draw[arr] (hc0) -- (hc0 |- out_c.north);
\draw[arr] (hc1) -- (hc1 |- out_c.north);
\draw[arr] (hc2) -- (hc2 |- out_c.north);
\node[font=\tiny, right=0.1cm of p2] {$\leftarrow$ task $k$};

\end{tikzpicture}
\caption{Three task routing architectures. (a)~Text-prefix (SH): task identity in input text, shared head. (b)~Multi-head (MH): task-specific heads, hard routing by task ID. (c)~MHP: task-specific transformer layers plus task-specific heads.}
\label{fig:architecture}
\end{figure*}

As our key novelty, we combine three ideas into a unified framework (Figure~\ref{fig:architecture}): (a)~a multi-task relevance model that jointly trains on six entity-pair tasks under a shared three-point relevance scale using both LoRA-adapted LLMs and fully fine-tuned cross-encoders, (b)~a systematic comparison of three task routing architectures---text-prefix routing, multi-head classification, and multi-head with private transformer layers---suggesting the encoder--decoder asymmetry described above, and (c)~a majority-vote ensemble using private-layer models that achieves the highest accuracy by exploiting the diversity induced by task-specific private layers.

Our experiments on 453K test examples yield three main results. First, private-layer routing (MHP) enables the strongest ensemble (89.96\%), surpassing the best individual model by +0.70\% and the text-prefix ensemble by +1.92\%. Second, an encoder--decoder asymmetry emerges in task routing: removing text prefixes barely affects cross-encoders ($-$1.02\%) but collapses decoder-only LLMs (Gemma-2-2b: $-$31.6\%), a failure that two private transformer layers (MHP) fully recover. Third, multi-task training lifts low-resource tasks by up to +14\%, with gains inversely correlated with per-task data volume.

Our contributions are: (a)~a unified multi-task relevance framework that jointly models six entity-pair relationship types under a shared three-point graded relevance scale, with a data derivation pipeline grounded in human-labeled seed data and validated through inter-annotator agreement studies; (b)~a systematic empirical comparison of LoRA-adapted decoder-only LLMs (2B and 8B parameters) versus fully fine-tuned cross-encoders (14M--396M parameters) across all six tasks; (c)~the first comparative study of three task routing architectures across both encoder and decoder model families, suggesting that private transformer layers are necessary for decoder-only LLMs but not for cross-encoders; and (d)~comprehensive ablation studies on task routing, dataset scale, class weighting, and model architecture, providing actionable guidance for practitioners deploying multi-task relevance models. While we evaluate this framework at Walmart, the approach is applicable to any e-commerce platform with human-labeled relevance data.

\section{Related Work}
\label{sec:related}

Our work sits at the intersection of multi-task learning, e-commerce relevance modeling, and task routing architectures. We review each area below, positioning our contributions relative to prior work.

\textbf{Multi-task learning} (MTL) enables knowledge transfer across related tasks~\cite{caruana1997multitask, ruder2017overview}. MT-DNN~\cite{liu2019mtdnn} showed that jointly training BERT~\cite{devlin2019bert} on multiple NLU tasks yields transferable representations. MTL architectures range from hard parameter sharing to soft parameter sharing and mixture-of-experts~\cite{crawshaw2020multitask, ma2018mmoe}. Tang et al.~\cite{tang2020ple} introduced Progressive Layered Extraction (PLE) with shared and task-specific expert networks, which is architecturally related to our MHP private layers. Standley et al.~\cite{standley2020tasks} showed that task groupings significantly impact performance and that negative transfer can occur with dissimilar tasks. Our work does not propose a new sharing mechanism. Instead, we fix the sharing strategy (hard parameter sharing) and systematically vary the task routing architecture---text-prefix, multi-head, and multi-head with private layers---across both encoder and decoder families, an asymmetry not studied in prior MTL work.

\textbf{E-commerce relevance modeling} has focused on query--product matching~\cite{aiello2016relevance, rao2020product}, with the ESCI dataset and KDD Cup 2022 shared task~\cite{reddy2022shopping} establishing a benchmark with graded relevance scales~\cite{jarvelin2002cumulated}. Rokon et al.~\cite{rokon2024enhancement} showed that LoRA-adapted Llama-2 achieves 89.43\% accuracy on query--ad relevance, outperforming GPT-4 (63\%) and BERT cross-encoders (86.27\%); our dataset seed comes from this work, and we extend it from a single query--ad task to six entity-pair tasks. Wang et al.~\cite{wang2024semantic} explored progressive training for semantic ad retrieval. Desai et al.~\cite{desai2026unified} introduced a unified supervision framework for retrieval by leveraging relevance and engagement signals. Our work does not focus on a single task. Instead, we extend query--ad relevance to a unified multi-task framework spanning six relationship types, enabling cross-task knowledge transfer that improves low-resource tasks by up to +14\%.

\textbf{LLMs for classification.} LoRA~\cite{hu2021lora} and QLoRA~\cite{dettmers2023qlora} enable parameter-efficient adaptation of billion-parameter models. Thomas et al.~\cite{thomas2024large} and Faggioli et al.~\cite{faggioli2023perspectives} showed that LLMs can serve as effective relevance judges. Prior work evaluates LLMs in isolation or against cross-encoders on single tasks. Our work differs by comparing LoRA-adapted Gemma~\cite{team2024gemma} and Llama-3~\cite{grattafiori2024llama3} against cross-encoders across six tasks simultaneously, showing that the two families make systematically different errors.

\textbf{Cross-encoders} process input pairs jointly for rich token-level interactions~\cite{nogueira2019passage, reimers2019sentencebert}. ModernBERT~\cite{warner2024modernbert} introduces rotary embeddings and flash attention; ELECTRA~\cite{clark2020electra} uses replaced-token-detection pretraining; TinyBERT~\cite{jiao2020tinybert} achieves competitive performance through distillation. Prior cross-encoder studies focus on single-task ranking or re-ranking. Our work extends them to multi-task classification and shows that their robustness to routing changes contrasts sharply with decoder-only LLMs.

\textbf{Ensemble methods} combine diverse models to improve robustness~\cite{dietterich2000ensemble}. In IR, reciprocal rank fusion~\cite{cormack2009reciprocal} outperforms individual rankers. Prior work typically ensembles models of the same architecture family. Our approach differs by combining architecturally diverse models---decoder-only LLMs and encoder-based cross-encoders---and showing that the private-layer routing architecture induces more diverse error patterns, yielding a stronger ensemble than text-prefix routing.

\section{Problem Formulation and Methods}
\label{sec:theory}

This section defines the problem, presents the multi-task learning objective, and describes the model architectures. We start with an intuitive overview and then introduce the formal notation.

\subsection{Problem Formulation}

The key idea of our problem formulation is to treat all six entity-pair relationship types as instances of a single three-class classification problem, differing only in which entity types form the input pair and how task identity is communicated to the model. Concretely, the e-commerce entity space comprises three entity types: queries $\mathcal{Q}$, ad items $\mathcal{A}$, and product types $\mathcal{P}$. Let $\mathcal{E} = \mathcal{Q} \cup \mathcal{A} \cup \mathcal{P}$ denote the full entity space. We define a set of six task types $\mathcal{T} = \{t_1, \ldots, t_6\}$, each specifying a particular entity-pair relationship, as detailed in Table~\ref{tab:tasks}. Given an entity pair and a task type, the model must predict one of three relevance levels: Irrelevant (0), Partially Relevant (1), or Relevant (2). Formally, the multi-task relevance problem is to learn a function:
\begin{equation}
    r: \mathcal{E} \times \mathcal{E} \times \mathcal{T} \rightarrow \mathcal{Y} = \{0, 1, 2\}
    \label{eq:relevance_function}
\end{equation}

\begin{table}[t]
\centering
\small
\caption{Task definitions for the six entity-pair relationship types. Totals are pre-filtering counts; see Table~\ref{tab:dataset} for the final dataset.}
\label{tab:tasks}
\begin{tabular}{@{}clllr@{}}
\toprule
\textbf{Task} & \textbf{Entity Pair} & \textbf{Notation} & \textbf{Example} & \textbf{Total} \\
\midrule
$t_1$ & Query--Ad Item & $(\mathcal{Q}, \mathcal{A})$ & ``hiking boots'' $\leftrightarrow$ Merrell Moab 3 & 800K \\
$t_2$ & Query--Query & $(\mathcal{Q}, \mathcal{Q})$ & ``hiking boots'' $\leftrightarrow$ ``outdoor footwear'' & 312K \\
$t_3$ & Query--Product Type & $(\mathcal{Q}, \mathcal{P})$ & ``hiking boots'' $\leftrightarrow$ Hiking Boots & 181K \\
$t_4$ & Ad Item--Product Type & $(\mathcal{A}, \mathcal{P})$ & Merrell Moab 3 $\leftrightarrow$ Hiking Boots & 1.1M \\
$t_5$ & Product Type--Product Type & $(\mathcal{P}, \mathcal{P})$ & Hiking Boots $\leftrightarrow$ Trail Shoes & 117K \\
$t_6$ & Ad Item--Ad Item & $(\mathcal{A}, \mathcal{A})$ & Merrell Moab 3 $\leftrightarrow$ Salomon X Ultra & 4.3M \\
\bottomrule
\end{tabular}
\end{table}

\subsection{Multi-Task Learning Formulation}

The key idea of multi-task learning in our setting is to train a single shared model on all six tasks simultaneously, so that knowledge from data-rich tasks (e.g., query--ad with 404K training examples) can transfer to data-scarce tasks (e.g., product-type--product-type with only 14K examples). The training loss is a weighted sum over all tasks, where each task's contribution is proportional to its share of the training data---this arises naturally from uniform sampling over the combined dataset. Formally, given $K = 6$ tasks, the joint objective is:
\begin{equation}
    \mathcal{L}_{\text{MTL}}(\theta) = \sum_{k=1}^{K} \frac{N_k}{N} \, \mathcal{L}_k(\theta)
    \label{eq:mtl_objective}
\end{equation}
where $\theta$ denotes the shared model parameters, $N_k$ is the number of examples for task $k$, and $N = \sum_k N_k$. We explore three mechanisms for communicating task identity to the shared model, ranging from text-prefix routing (Section~\ref{sec:routing}) with purely shared parameters, to multi-head architectures with task-specific classification heads or private transformer layers.

Each task uses categorical cross-entropy as the per-task loss. In plain terms, the loss penalizes the model for assigning low probability to the correct relevance class:
\begin{equation}
    \mathcal{L}_k(\theta) = -\frac{1}{N_k} \sum_{i=1}^{N_k} \sum_{c=0}^{2} y_{i,c}^{(k)} \log p_{i,c}^{(k)}(\theta)
    \label{eq:ce_loss}
\end{equation}
where $y_{i,c}^{(k)}$ is a binary indicator of whether class $c$ is the correct label for example $i$ in task $k$, and $p_{i,c}^{(k)}(\theta)$ is the predicted probability.

Because the ``Partially Relevant'' class is harder to classify than the two extremes, we also experiment with weighted cross-entropy that upweights the middle class:
\begin{equation}
    \mathcal{L}_k^{w}(\theta) = -\frac{1}{N_k} \sum_{i=1}^{N_k} \sum_{c=0}^{2} w_c \cdot y_{i,c}^{(k)} \log p_{i,c}^{(k)}(\theta)
    \label{eq:weighted_ce}
\end{equation}
where $\mathbf{w} = [w_0, w_1, w_2]$ are class weights. We experiment with $\mathbf{w} \in \{[1, 1, 1], [1, 1.25, 1], [1, 1.5, 1]\}$ to study the trade-off between overall accuracy and per-class balance (Section~\ref{sec:ablations}).

All models are optimized using AdamW~\cite{loshchilov2019adamw} with decoupled weight decay.

\subsection{Low-Rank Adaptation (LoRA)}

LoRA~\cite{hu2021lora} keeps the pre-trained weight matrix $W \in \mathbb{R}^{d \times h}$ frozen and injects a low-rank update $\Delta W = BA$ with $B \in \mathbb{R}^{d \times r}$, $A \in \mathbb{R}^{r \times h}$, and $r \ll \min(d, h)$:
\begin{equation}
    \hat{W} = W + BA
    \label{eq:lora}
\end{equation}
With rank $r = 16$, this yields ${\sim}$2M trainable parameters for Gemma-2-2b (0.1\% of total) and ${\sim}$4M for Llama-3-8b (0.05\%).

\subsection{Cross-Encoder Scoring}

The key idea of cross-encoder scoring is to feed both entities as a single concatenated sequence through a bidirectional encoder, so that every token in one entity can attend to every token in the other. This enables rich token-level interactions that are not possible with bi-encoder approaches. Formally, for a cross-encoder with parameters $\psi$, the relevance distribution is:
\begin{equation}
    p(y \mid e_i, e_j; \psi) = \text{softmax}\left(W_c \cdot h_{\text{[CLS]}}(e_i, e_j; \psi)\right)
    \label{eq:cross_encoder}
\end{equation}
where $h_{\text{[CLS]}}$ is the pooled representation from the encoder's classification token and $W_c \in \mathbb{R}^{3 \times d_h}$ is the classification head. Unlike LoRA-adapted LLMs that process input autoregressively, cross-encoders compute bidirectional attention over the concatenated input.

\subsection{Ensemble via Majority Voting}

The key idea of our ensemble is simple: if two out of three architecturally diverse models agree on a prediction, that prediction is likely correct. When all three disagree, we fall back to the model with the highest confidence. Formally, given $M = 3$ models with predictions $\{\hat{y}_1, \hat{y}_2, \hat{y}_3\}$ and softmax vectors $\{p_1, p_2, p_3\}$:
\begin{equation}
    \hat{y}_{\text{ens}} = \begin{cases}
        \text{mode}(\hat{y}_1, \hat{y}_2, \hat{y}_3) & \text{if at least 2 of 3 models agree} \\
        \arg\max_c \frac{1}{M} \sum_{m=1}^{M} p_m(c) & \text{otherwise (all three disagree)}
    \end{cases}
    \label{eq:ensemble}
\end{equation}

The effectiveness of majority voting depends on the \emph{diversity} of errors across constituent models~\cite{dietterich2000ensemble}. In our setting, the three models exhibit substantial disagreement: Llama-3-8b and ModernBERT disagree on 14.2\% of test examples, with disagreement concentrated on abstract tasks (T3: 26.8\%) and minimal on lexically-rich tasks (T6: 4.1\%). This architectural diversity---decoder-only LLMs and encoder-based cross-encoders making systematically different errors---is the property that enables the ensemble to correct individual model failures. We chose majority voting over learned combinations (e.g., stacking) because it requires no calibration set, is transparent---each model's vote carries equal weight---and avoids overfitting when the constituent models are few ($M=3$).

\subsection{Task Routing Architectures}
\label{sec:routing}

The key idea of task routing is to answer a deceptively simple question: how should a shared multi-task model know which task it is solving for a given input? We formalize and compare three architectures, each encoding task identity differently.

\noindent\textbf{Text-Prefix Routing (Baseline).} The simplest approach prepends a task-descriptive string to the input, letting the model learn to condition on the prefix tokens. For task $t_k$ with entity pair $(e_i, e_j)$, the formatted input is:
\begin{equation}
    x_k = \text{concat}\left(\tau_k, \; e_i, \; \texttt{[SEP]}, \; e_j\right)
    \label{eq:prefix_routing}
\end{equation}
where $\tau_k$ is a task-specific prefix string (e.g., ``Task Query and Product Title Relevance:''). A single shared classification head $W_c \in \mathbb{R}^{|\mathcal{Y}| \times d_h}$ maps the pooled representation to class logits. Task-specific behavior emerges implicitly through the model learning to condition on the prefix tokens. This approach adds zero additional parameters but conflates task routing with input representation.

\noindent\textbf{Multi-Head Classification (MH).}\label{sec:mh} The key idea of MH is to decouple task routing from input encoding by replacing the shared classification head with $K$ task-specific heads, each a simple linear projection. The input no longer contains a task prefix:
\begin{equation}
    x = \text{concat}\left(e_i, \; \texttt{[SEP]}, \; e_j\right)
    \label{eq:multihead_input}
\end{equation}
Given the shared encoder output $h = f_\theta(x)$ (where $h = h_\text{[CLS]}$ for cross-encoders or $h = h_\text{last}$ for causal LLMs), the prediction for task $t_k$ is:
\begin{equation}
    p(y \mid e_i, e_j, t_k) = \text{softmax}\left(W_c^{(k)} \cdot h\right), \quad W_c^{(k)} \in \mathbb{R}^{|\mathcal{Y}| \times d_h}
    \label{eq:multihead}
\end{equation}
where the task index $k$ deterministically selects the appropriate head (hard routing). This adds $K \times |\mathcal{Y}| \times d_h$ parameters---approximately 18K for our cross-encoder configuration ($K=6$, $|\mathcal{Y}|=3$, $d_h=1024$), a negligible overhead.

\noindent\textbf{Multi-Head with Private Layers (MHP).}\label{sec:mhp} The key idea of MHP is to give each task its own ``private'' transformer layers on top of the shared backbone, providing richer task-specific representation capacity than a single linear projection. MHP inserts $L$ private transformer encoder layers between the shared encoder and each task head. We define a pooling function $\text{pool}(\cdot)$ that extracts the [CLS] token representation for cross-encoders or the last-token representation for causal LLMs:
\begin{equation}
    h_k = g_{\phi_k}^{(1:L)}(f_\theta(x)), \quad p(y \mid e_i, e_j, t_k) = \text{softmax}\left(W_c^{(k)} \cdot \text{pool}(h_k)\right)
    \label{eq:private_layers}
\end{equation}
where $g_{\phi_k}^{(1:L)}$ denotes a stack of $L$ transformer encoder layers with parameters $\phi_k$ private to task $k$, operating on the full sequence of hidden states from the shared backbone.

Each private layer follows the standard transformer encoder architecture with multi-head self-attention and a feed-forward network, using a reduced FFN dimension to control parameter overhead. With $L=2$ private layers, MHP adds approximately 24M parameters for cross-encoders (hidden size 1024, 8 attention heads, FFN dimension 1024) or 402M for LLMs (hidden size 2048/3072, 8 heads, FFN dimension 2048), distributed across the $K=6$ task stacks.

\textbf{Comparison with Mixture-of-Experts.} Unlike soft-routing MoE architectures~\cite{ma2018mmoe} where a gating network learns to blend expert outputs, our approach uses \emph{hard deterministic routing}: the task identity is known at both training and inference time, so no learned gating is required. This eliminates load-balancing losses, reduces routing overhead, and ensures that each task's private layers are always fully used.

\section{Dataset Construction}
\label{sec:data}

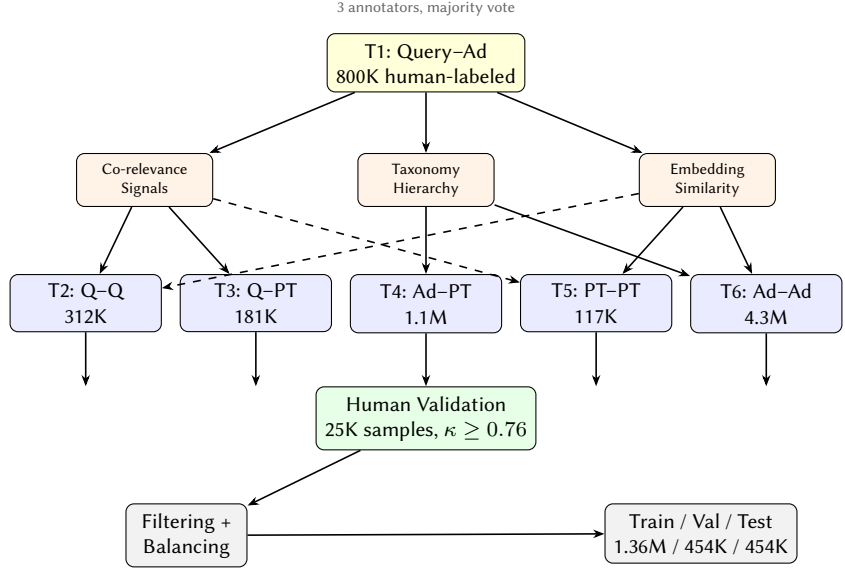
\begin{figure*}[t]
\centering
\begin{tikzpicture}[
    node distance=0.4cm and 0.6cm,
    box/.style={rectangle, draw, rounded corners=3pt, minimum height=0.7cm, align=center, font=\scriptsize},
    seed/.style={box, fill=yellow!15, minimum width=2.2cm},
    derive/.style={box, fill=blue!8, minimum width=2.0cm},
    signal/.style={box, fill=orange!10, minimum width=1.8cm, font=\tiny},
    valid/.style={box, fill=green!10, minimum width=2.0cm},
    data/.style={box, fill=gray!10, minimum width=1.6cm},
    arr/.style={-{Stealth[length=1.5mm]}, semithick},
    darr/.style={-{Stealth[length=1.5mm]}, semithick, dashed},
    lbl/.style={font=\tiny, text=black!60},
]

\node[seed] (seed) {T1: Query--Ad\\800K human-labeled};
\node[lbl, above=0.1cm of seed] {3 annotators, majority vote};

\node[signal, below left=0.8cm and 1.5cm of seed] (cosig) {Co-relevance\\Signals};
\node[signal, below=0.8cm of seed] (taxo) {Taxonomy\\Hierarchy};
\node[signal, below right=0.8cm and 1.5cm of seed] (embed) {Embedding\\Similarity};

\draw[arr] (seed) -- (cosig);
\draw[arr] (seed) -- (taxo);
\draw[arr] (seed) -- (embed);

\node[derive, below=2.4cm of seed, xshift=-4.5cm] (t2) {T2: Q--Q\\312K};
\node[derive, below=2.4cm of seed, xshift=-2.25cm] (t3) {T3: Q--PT\\181K};
\node[derive, below=2.4cm of seed, xshift=0cm] (t4) {T4: Ad--PT\\1.1M};
\node[derive, below=2.4cm of seed, xshift=2.25cm] (t5) {T5: PT--PT\\117K};
\node[derive, below=2.4cm of seed, xshift=4.5cm] (t6) {T6: Ad--Ad\\4.3M};

\draw[arr] (cosig) -- (t2);
\draw[arr] (cosig) -- (t3);
\draw[arr] (taxo) -- (t4);
\draw[arr] (embed) -- (t5);
\draw[arr] (taxo) -- (t6);
\draw[arr] (embed) -- (t6);
\draw[darr] (embed) -- (t2);
\draw[darr] (cosig) -- (t5);

\node[valid, below=0.7cm of t4] (hval) {Human Validation\\25K samples, $\kappa \geq 0.76$};
\draw[arr] (t2) -- (t2 |- hval.north);
\draw[arr] (t3) -- (t3 |- hval.north);
\draw[arr] (t4) -- (hval);
\draw[arr] (t5) -- (t5 |- hval.north);
\draw[arr] (t6) -- (t6 |- hval.north);

\node[data, below left=0.7cm and 0.9cm of hval] (filter) {Filtering +\\Balancing};
\node[data, below right=0.7cm and 0.9cm of hval] (split) {Train / Val / Test\\1.36M / 454K / 454K};
\draw[arr] (hval) -- (filter);
\draw[arr] (filter) -- (split);

\end{tikzpicture}
\caption{Dataset construction pipeline. Human-labeled query--ad pairs (T1) seed five derived tasks via co-relevance, taxonomy, and embedding signals. All derived tasks undergo human validation ($\kappa \geq 0.76$), filtering, and balancing to produce the final 2.27M-example dataset.}
\label{fig:data_pipeline}
\end{figure*}

The key idea behind our dataset construction is to start from a high-quality human-labeled seed (query--ad pairs) and systematically derive the remaining five tasks using co-relevance signals, taxonomy hierarchies, and embedding similarity, with human validation at each step. Figure~\ref{fig:data_pipeline} illustrates the full pipeline.

\noindent\textbf{Human-Labeled Seed Data (Task 1).} The foundation of our dataset is approximately 800K query--ad item pairs from Walmart's sponsored search logs. Each pair was scored by three independent annotators on a three-point scale (0: Irrelevant, 1: Partially Relevant, 2: Relevant), with the majority vote determining the final label~\cite{rokon2024enhancement}.

\noindent\textbf{Systematic Derivation of Tasks 2--6.} Building upon the human-labeled query--ad pairs, we derive training data for the remaining five tasks through principled algorithms that use co-relevance signals, taxonomy hierarchies, and embedding similarity:

\textbf{Task 2 (Query--Query):} For each query $q_1$ with known ad relevance scores, we identify other queries $q_2$ that share highly relevant ads. We compute query-pair labels using both co-relevance scores and cosine similarity from a text embedding model. Pairs with co-relevance $\geq 1.5$ and similarity $\geq 0.65$ receive label 2; pairs with moderate co-relevance ($0.5$--$1.5$) and similarity ($0.4$--$0.65$) receive label 1; pairs with low co-relevance ($< 0.5$) and similarity ($< 0.35$) receive label 0. This yields 312K query--query pairs.

\textbf{Task 3 (Query--Product Type):} We aggregate query--ad relevance scores at the product-type level and incorporate product-type prediction API scores. Queries with high average relevance to ads of a given product type ($\geq 0.8$) or strong API prediction scores receive label 2; moderate cases receive label 1; low-relevance pairs receive label 0. This yields 181K pairs.

\textbf{Task 4 (Ad Item--Product Type):} We use the product taxonomy hierarchy. Exact product-type matches receive label 2; matches within the same taxonomy branch (top-3 or top-2 level) receive label 1; mismatches at the top-1 level receive label 0. This yields 1.1M pairs.

\textbf{Task 5 (Product Type--Product Type):} We identify product-type pairs that co-occur in query--ad contexts and compute embedding similarity on representative ad titles. Pairs with high co-relevance and similarity ($\geq 0.7$) receive label 2; moderate similarity ($0.4$--$0.7$) receives label 1; low similarity ($< 0.2$) receives label 0. This yields 117K pairs.

\textbf{Task 6 (Ad Item--Ad Item):} We combine product-type matching, taxonomy hierarchy alignment, and title embedding similarity. Same product-type pairs with matching relevance labels receive label 2; cross-type pairs with taxonomy alignment and moderate similarity receive label 1; pairs with taxonomy mismatch and low similarity ($< 0.3$) receive label 0. This yields 4.3M pairs.

\textbf{Post-processing:} We apply similarity-based filtering to remove ambiguous pairs (e.g., removing label-2 pairs with similarity $< 0.7$ and label-0 pairs with similarity $> 0.2$) and perform class-balanced sampling to ensure approximately equal representation of the three relevance levels. The derivation thresholds above were tuned on a held-out validation set of 2,000 pairs per task to maximize derived--human agreement; a sensitivity analysis showed $\pm$0.05 threshold variation changes accuracy by $<$1\%.

\noindent\textbf{Human Validation of Derived Labels.} To validate the quality of algorithmically derived labels, we conducted a human evaluation study. We sampled approximately 5,000 examples per derived task (T2--T6), totaling 25,000 pairs. Each pair was independently labeled by two trained annotators on the same three-point scale. We measured inter-annotator agreement using Cohen's $\kappa$~\cite{cohen1960kappa} and agreement between annotators and derived labels. Results are presented in Table~\ref{tab:validation}.

\begin{table}[t]
\centering
\small
\caption{Human validation of derived labels. Cohen's $\kappa$ measures inter-annotator agreement. All $\kappa$ values have 95\% CI width $< \pm$0.03.}
\label{tab:validation}
\begin{tabular}{@{}lcccc@{}}
\toprule
\textbf{Task} & \textbf{Samples} & \textbf{Cohen's $\kappa$} & \textbf{Human--Human} & \textbf{Derived--Human} \\
\midrule
T2 (Query--Query) & 5,000 & 0.82 & 89.3\% & 88.1\% \\
T3 (Query--PT) & 5,000 & 0.76 & 83.7\% & 82.4\% \\
T4 (Ad--PT) & 5,000 & 0.84 & 91.2\% & 90.5\% \\
T5 (PT--PT) & 5,000 & 0.79 & 86.5\% & 85.8\% \\
T6 (Ad--Ad) & 5,000 & 0.85 & 92.1\% & 91.3\% \\
\bottomrule
\end{tabular}
\end{table}

The lower agreement for Task~3 (Query--PT, $\kappa = 0.76$) reflects the inherent ambiguity of query-to-product-type relevance---a pattern that is also reflected in model performance, as Task~3 proves to be the most challenging for all architectures (Section~\ref{sec:experiments}).

\noindent\textbf{Final Dataset Statistics.} Table~\ref{tab:dataset} summarizes the final dataset (V2, large version). The dataset comprises 1.36M training, 454K validation, and 454K test examples across all six tasks, with approximately balanced label distributions.

\begin{table}[t]
\centering
\caption{Dataset statistics (V2). Label distribution in training set: Irrelevant 454K (33.4\%), Partially Relevant 454K (33.3\%), Relevant 453K (33.3\%).}
\label{tab:dataset}
\small
\begin{tabular}{@{}llrrrr@{}}
\toprule
\textbf{Task} & \textbf{Entity Pair} & \textbf{Train} & \textbf{Val} & \textbf{Test} & \textbf{\% Train} \\
\midrule
T1 & Query--Ad & 404,146 & 134,715 & 134,716 & 29.71\% \\
T2 & Query--Query & 142,725 & 47,575 & 47,576 & 10.49\% \\
T3 & Query--PT & 33,670 & 11,223 & 11,224 & 2.47\% \\
T4 & Ad--PT & 388,128 & 129,376 & 129,377 & 28.53\% \\
T5 & PT--PT & 13,798 & 4,599 & 4,600 & 1.01\% \\
T6 & Ad--Ad & 378,052 & 126,017 & 126,018 & 27.79\% \\
\midrule
\textbf{Total} & & \textbf{1,360,519} & \textbf{453,505} & \textbf{453,511} & \\
\bottomrule
\end{tabular}
\end{table}

\section{Model Architectures and Training}
\label{sec:models}

We evaluate two families of models: (a)~LoRA-adapted decoder-only LLMs, which are parameter-efficient but process input autoregressively, and (b)~fully fine-tuned cross-encoders, which are smaller but compute bidirectional attention. Figure~\ref{fig:system_overview} shows the end-to-end pipeline from data construction through training to ensemble deployment. This section details the specific models, training configurations, and multi-head variants for each family.

\noindent\textbf{LoRA-Adapted Large Language Models.} We employ decoder-only transformers with a three-class linear classification head appended after the final hidden state. LoRA is applied to all attention projections ($q$, $k$, $v$, $o$) and feed-forward layers ($gate$, $up$, $down$) of each transformer block.

\textbf{Gemma-2-2b}~\cite{team2024gemma}: A 2-billion parameter decoder-only model from Google. With LoRA rank $r = 16$, approximately 2M parameters (0.1\% of total) are trainable.

\textbf{Llama-3-8b}~\cite{grattafiori2024llama3}: An 8-billion parameter decoder-only model from Meta. With LoRA rank $r = 16$, approximately 4M parameters (0.05\% of total) are trainable.

The input is formatted as a task-prefixed prompt: \texttt{``Task Query and Product Title Relevance: \{$q_1$\} [SEP] \{$q_2$\}''}. The classification probability is computed as:
\begin{equation}
    p(y \mid e_i, e_j, t) = \text{softmax}\left(W_c \cdot h_{\text{last}}\left(\text{prompt}(e_i, e_j, t); \theta + \Delta\theta\right)\right)
    \label{eq:lora_classification}
\end{equation}
where $h_{\text{last}}$ is the hidden state at the last token position and $\Delta\theta = BA$ represents the LoRA adaptation.

\textbf{Training Setup:} AdamW optimizer, learning rate $1 \times 10^{-5}$, batch size 16, 3 epochs, 2$\times$ NVIDIA A100-SXM4-40GB GPUs. Tokenizer padding is set to the left with \texttt{pad\_token = eos\_token} and maximum sequence length of 64 tokens. With this limit, approximately 3.2\% of LLM inputs are truncated; truncation primarily affects Task~6 (Ad--Ad) where product titles can be long.

\noindent\textbf{Cross-Encoder Models.} Cross-encoder models process the concatenated input pair through a bidirectional encoder, with the [CLS] token representation passed through a classification head~(Eq.~\ref{eq:cross_encoder}). We evaluate five cross-encoder architectures: (a)~\textbf{ModernBERT-large} (396M params)~\cite{warner2024modernbert}, initialized from \texttt{reranker-ModernBERT-large-gooaq-bce}, featuring rotary positional embeddings and flash attention; (b)~\textbf{ELECTRA-base} (109M params)~\cite{clark2020electra}, initialized from \texttt{cross-encoder/qnli-electra-base}, with replaced-token-detection pretraining; (c)~\textbf{TinyBERT} (14M params)~\cite{jiao2020tinybert}, initialized from \texttt{cross-encoder/stsb-TinyBERT-L4}, a distilled compact model; (d)~\textbf{RoBERTa-large} (355M params)~\cite{liu2019roberta}, initialized from \texttt{cross-encoder/stsb-roberta-large}; and (e)~\textbf{BGE-reranker} (278M params)~\cite{xiao2024bge}, initialized from \texttt{BAAI/bge-reranker-base}.

\textbf{Training Setup:} Full fine-tuning (all parameters trainable), AdamW optimizer, learning rate $2 \times 10^{-5}$, batch size 16, 10 epochs, class weights $[1.0, 1.25, 1.0]$, maximum sequence length 128 tokens, 1$\times$ NVIDIA A100-SXM4-40GB GPU.

\noindent\textbf{Multi-Head Variants (MH and MHP).} For each model family, we additionally train multi-head (MH) and multi-head with private layers (MHP) variants, as described in Section~\ref{sec:routing}. The text prefix is removed from the input and task identity is provided as an integer ID that selects the appropriate classification head.

\textbf{Cross-encoder MH/MHP:} Trained with full fine-tuning in fp32, AdamW optimizer, learning rate $2 \times 10^{-5}$, batch size 64, up to 50 epochs with early stopping (patience~5 epochs on validation accuracy), class weights $[1.0, 1.25, 1.0]$, maximum sequence length 64 tokens. MHP uses $L=2$ private transformer layers with 8 attention heads and FFN dimension 1024.

\textbf{LLM MH/MHP:} Trained with LoRA ($r=16$, $\alpha=16$, dropout 0.1) in bfloat16 mixed precision using \texttt{torch.amp.autocast}. AdamW optimizer, learning rate $1 \times 10^{-5}$, batch size 16, up to 4 epochs with early stopping (patience~5). MHP uses $L=2$ private transformer layers with 8 attention heads and reduced FFN dimension ($\min(d_h, 2048)$). Models are compiled with \texttt{torch.compile} for training acceleration.

The LoRA \texttt{modules\_to\_save} list is extended to include the task-specific heads (MH) or both private layers and task-specific heads (MHP), ensuring these parameters are saved alongside the LoRA adapter weights.

\noindent\textbf{Majority-Vote Ensemble.} We explore combining the best model from each family---Gemma-2-2b, Llama-3-8b, and ModernBERT---via majority voting (Eq.~\ref{eq:ensemble}). We evaluate two ensemble configurations: one using text-prefix routing models (SH Ensemble) and one using multi-head with private layers models (MHP Ensemble).

\section{Experiments and Results}
\label{sec:experiments}

We present our results in four parts: (1)~main results comparing baselines, task routing architectures, and ensembles; (2)~per-class analysis showing the difficulty of partial relevance; (3)~per-task breakdown showing complementary strengths; and (4)~ablation studies isolating key design choices (Section~\ref{sec:ablations}). We begin with the experimental setup.

\noindent\textbf{Experimental Setup.}
\textbf{Hardware.} Training was conducted on NVIDIA A100-SXM4-40GB GPUs; inference evaluation on Tesla V100-SXM2-32GB GPUs.

\textbf{Metrics.} We report both sample-weighted overall accuracy and macro-averaged per-task accuracy (unweighted mean across all six tasks), along with per-class precision, recall, and F1-score. All results are on the held-out test set comprising 453,511 examples. We assess statistical significance using a paired bootstrap test~(10,000 resamples); at this sample size, all differences $\geq$0.5\% are significant at $p < 0.01$.

\textbf{Training regime differences.} The LoRA-adapted LLMs and cross-encoders were trained under their respective best-practice configurations rather than identical hyperparameters. Key differences include: sequence length (64 vs.\ 128 tokens), training epochs (3 vs.\ 10), class weighting ($[1, 1, 1]$ vs.\ $[1, 1.25, 1]$), and adaptation method (LoRA rank 16, $\sim$0.1\% trainable vs.\ full fine-tuning). The cross-family comparison (LLM vs.\ cross-encoder) is therefore observational rather than a controlled ablation; the routing ablation within each family (Section~\ref{sec:routing_ablation}) provides the controlled comparison by varying only the routing architecture while holding the backbone, optimizer, and data constant.

\noindent\textbf{Baselines.} Off-the-shelf models without task-specific fine-tuning perform poorly on our multi-task benchmark, suggesting that domain-specific training is necessary. We compare against two baseline categories. \textbf{Off-the-shelf models} (zero-shot or few-shot): GPT-4~\cite{achiam2023gpt4} with 5-shot prompting (task description plus one labeled example per class, majority vote over 3~API calls, temperature 0.3) achieves only 67.2\% accuracy, consistent with Rokon et al.~\cite{rokon2024enhancement} who reported 63\% on single-task query--ad relevance. Zero-shot Llama-3-8b reaches only 61.3\%. These models lack calibration for the three-point relevance scale. \textbf{Single-task baselines} (same architectures, trained per-task individually): for each of the six tasks, we train a dedicated model using the same hyperparameters as its multi-task counterpart (LoRA rank~16, lr~$1{\times}10^{-5}$, 3~epochs for LLMs; full fine-tuning, lr~$2{\times}10^{-5}$, 10~epochs for cross-encoders). The ``Overall'' and ``Macro'' columns aggregate the six single-task models' predictions---each task's test predictions come from its own dedicated model, combined and weighted by test set size (Overall) or averaged equally (Macro). Single-task models suffer on low-resource tasks because they cannot use cross-task knowledge---the cross-encoder single-task model achieves 95.82\% on Ad--Ad but only 57.38\% on Query--PT. Single-task deployment also requires $6\times$ the training and serving infrastructure. All baseline results appear in Table~\ref{tab:main_results}.

\noindent\textbf{Main Results.}\label{sec:main_results} The MHP Ensemble achieves the highest accuracy (89.96\%) across all configurations, outperforming the best single model (ModernBERT SH: 89.26\%) by +0.70\% and the text-prefix ensemble by +1.92\%. Table~\ref{tab:main_results} consolidates all results across baselines, multi-task models with text-prefix routing (SH), multi-head (MH), multi-head with private layers (MHP), and ensemble configurations.

\begin{table*}[t]
\centering
\small
\caption{Main results on 453,511 test examples across six entity-pair tasks. Per-task accuracy (\%). \textbf{Overall} = sample-weighted accuracy; \textbf{Macro} = unweighted mean of per-task accuracies. Best result per metric in \textbf{bold}. $^\dagger$Gemma MH exhibits training collapse (Section~\ref{sec:routing_ablation}). Labels sourced from human annotation (T1) and validated derivation (T2--T6, $\kappa \geq 0.76$). All pairwise differences $\geq$0.5\% are statistically significant ($p < 0.01$, paired bootstrap, 10K resamples).}
\label{tab:main_results}
\resizebox{\textwidth}{!}{%
\begin{tabular}{@{}llccrrrrrrrr@{}}
\toprule
\textbf{Category} & \textbf{Model} & \textbf{Routing} & \textbf{Params} & \textbf{Overall} & \textbf{Macro} & \textbf{T1} & \textbf{T2} & \textbf{T3} & \textbf{T4} & \textbf{T5} & \textbf{T6} \\
\midrule
\multirow{2}{*}{Off-the-shelf} & GPT-4 (few-shot) & -- & $\sim$1.8T & 67.20 & 66.22 & 72.10 & 69.40 & 58.30 & 65.80 & 61.20 & 70.50 \\
 & Llama-3-8b (zero) & -- & 8B & 61.30 & 60.18 & 65.40 & 63.50 & 52.80 & 59.70 & 55.40 & 64.30 \\
\midrule
\multirow{3}{*}{Single-task} & Gemma-2-2b & -- & 2B & 83.42 & 81.04 & 86.71 & 85.23 & 72.15 & 84.38 & 69.82 & 87.94 \\
 & Llama-3-8b & -- & 8B & 84.89 & 82.40 & 87.45 & 86.12 & 74.83 & 85.91 & 71.47 & 88.63 \\
 & ModernBERT & -- & 396M & 85.73 & 82.20 & 84.21 & 91.45 & 57.38 & 86.12 & 78.23 & 95.82 \\
\midrule
\multirow{4}{*}{\shortstack[l]{Multi-task\\(Text-Prefix)}} & Gemma-2-2b & SH & 2B & 87.14 & 83.56 & 81.96 & 92.45 & 62.91 & 85.83 & 83.89 & 94.30 \\
 & Llama-3-8b & SH & 8B & 87.88 & 84.63 & 83.02 & 93.25 & 64.67 & 86.08 & 85.74 & 95.02 \\
 & ModernBERT & SH & 396M & 89.26 & 85.55 & 82.58 & 93.90 & 62.30 & 88.56 & 88.13 & 97.81 \\
 & ELECTRA & SH & 109M & 88.38 & 84.73 & 82.14 & 93.28 & 63.58 & 87.36 & 85.80 & 96.57 \\
\midrule
\multirow{3}{*}{\shortstack[l]{Multi-task\\(Multi-Head)}} & Gemma-2-2b$^\dagger$ & MH & 2B & 55.59 & 46.43 & 47.36 & 33.16 & 33.14 & 44.81 & 33.41 & 86.71 \\
 & Llama-3-8b & MH & 8B & 80.96 & 79.97 & 77.90 & 92.66 & 62.22 & 72.27 & 84.50 & 90.28 \\
 & ModernBERT & MH & 396M & 88.24 & 83.88 & 81.43 & 92.98 & 60.14 & 87.76 & 84.11 & 96.88 \\
\midrule
\multirow{3}{*}{\shortstack[l]{Multi-task\\(MH+Private)}} & Gemma-2-2b & MHP & 2B & 88.04 & 83.76 & 82.49 & 92.33 & 61.18 & 86.99 & 83.59 & 95.99 \\
 & Llama-3-8b & MHP & 8B & 88.27 & 84.25 & 82.69 & 92.73 & 61.74 & 87.38 & 85.04 & 95.93 \\
 & ModernBERT & MHP & 396M & 88.31 & 83.95 & 81.50 & 93.00 & 60.20 & 87.86 & 84.20 & 96.94 \\
\midrule
\multirow{2}{*}{Ensemble} & SH Ensemble & SH & -- & 88.04 & 84.46 & 82.46 & 92.90 & 63.41 & 86.54 & 85.46 & 96.02 \\
 & MHP Ensemble & MHP & -- & \textbf{89.96} & \textbf{86.19} & \textbf{84.64} & \textbf{94.38} & \textbf{64.21} & \textbf{89.07} & \textbf{87.61} & \textbf{97.26} \\
\bottomrule
\end{tabular}%
}
\end{table*}

Multi-task training yields large gains over single-task baselines, especially on low-resource tasks. Compared to single-task Gemma-2-2b, the multi-task version (SH) improves by +3.72\% overall (83.42\% $\to$ 87.14\%), with the largest gains on low-resource tasks: +14.07\% on T5 (PT--PT, from 69.82\% to 83.89\%) and +7.22\% on T2 (Q--Q, from 85.23\% to 92.45\%). These gains are consistent with positive transfer from data-rich to data-scarce tasks.

\begin{figure}[t]
\centering
\small
\begin{tikzpicture}
\begin{axis}[
    ybar,
    bar width=3.5pt,
    width=\columnwidth,
    height=5cm,
    ylabel={Accuracy (\%)},
    symbolic x coords={T1,T2,T3,T4,T5,T6},
    xtick=data,
    ymin=55, ymax=100,
    legend style={at={(0.5,1.02)}, anchor=south, legend columns=2, font=\scriptsize},
    every axis plot/.append style={fill opacity=0.85},
    enlarge x limits=0.12,
    grid=major,
    grid style={dashed, gray!30},
]
\addplot coordinates {(T1,86.71) (T2,85.23) (T3,72.15) (T4,84.38) (T5,69.82) (T6,87.94)};
\addplot coordinates {(T1,82.58) (T2,93.90) (T3,62.30) (T4,88.56) (T5,88.13) (T6,97.81)};
\addplot coordinates {(T1,82.69) (T2,92.73) (T3,61.74) (T4,87.38) (T5,85.04) (T6,95.93)};
\addplot coordinates {(T1,81.50) (T2,93.00) (T3,60.20) (T4,87.86) (T5,84.20) (T6,96.94)};
\addplot coordinates {(T1,84.64) (T2,94.38) (T3,64.21) (T4,89.07) (T5,87.61) (T6,97.26)};
\legend{Single-task Gemma, ModernBERT SH, Llama MHP, ModernBERT MHP, MHP Ensemble}
\end{axis}
\end{tikzpicture}
\caption{Per-task accuracy on 453K test examples. MHP Ensemble achieves the best accuracy on all six tasks. Cross-encoders excel on lexically-rich tasks (T2, T6) but struggle on abstract tasks (T3).}
\label{fig:per_task}
\end{figure}
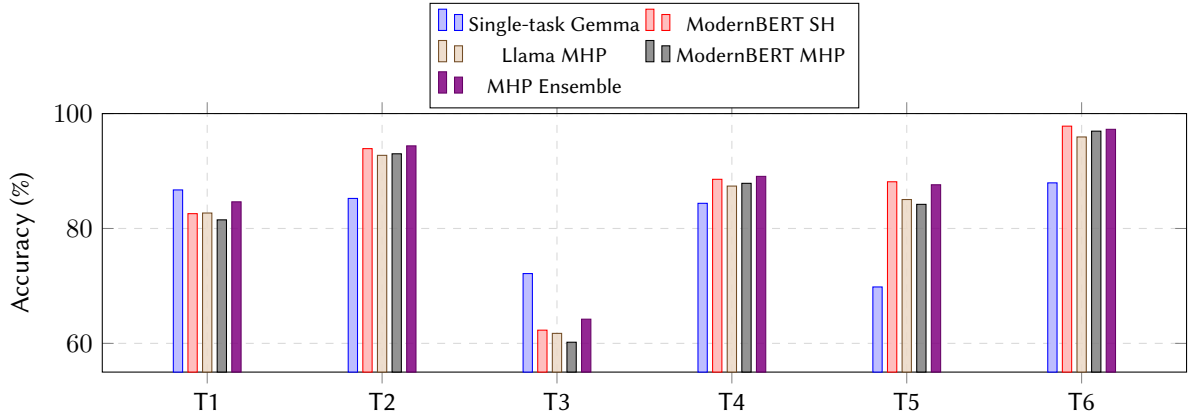

The two architecture families show complementary strengths (Figure~\ref{fig:per_task}). LoRA-adapted LLMs show uniform performance across all tasks with text-prefix routing, while cross-encoders show high variance (ModernBERT SH range: 62.30\%--97.81\%). Although ModernBERT SH achieves the highest single-model accuracy (89.26\%), its strength is concentrated on lexically-rich tasks (T6 Ad--Ad: 97.81\%, T2 Q--Q: 93.90\%) while it struggles on abstract tasks (T3 Q--PT: 62.30\%).

The SH Ensemble (88.04\%) underperforms individual ModernBERT SH (89.26\%) on sample-weighted accuracy because the two LLM members (Gemma: 87.14\%, Llama: 87.88\%) are weaker than ModernBERT on the dominant tasks (T1, T4, T6 together constitute 86\% of test data). In majority voting, two weaker models can outvote a stronger one on these high-traffic tasks. The MHP Ensemble avoids this because MHP training narrows the inter-model gap, achieving 89.96\% overall and 86.19\% macro-averaged accuracy---the best on both metrics---surpassing the SH Ensemble by +1.92\%. This improvement suggests that the multi-head architecture with private layers induces more diverse per-model error patterns, yielding a stronger ensemble. The task routing comparison, our central ablation, is analyzed in detail in Section~\ref{sec:routing_ablation}.

\noindent\textbf{Per-Class Analysis.}\label{sec:per_class} The ``Partially Relevant'' class is consistently the hardest across all architectures, with F1 scores 3--8\% lower than the ``Irrelevant'' and ``Relevant'' classes (Table~\ref{tab:per_class}). This is expected: partial relevance is inherently subjective, and even human annotators show lower agreement on this class ($\kappa = 0.76$--$0.85$ across tasks). MHP models achieve comparable per-class distributions to their SH counterparts, suggesting that private layers recover class-level discrimination without degradation.

\begin{table}[t]
\centering
\small
\caption{Per-class precision, recall, and F1 for SH and MHP models on 453K test examples. ``Partially Relevant'' is hardest for all architectures.}
\label{tab:per_class}
\begin{tabular}{@{}lllcccc@{}}
\toprule
\textbf{Model} & \textbf{Rt.} & \textbf{Class} & \textbf{Acc} & \textbf{Prec} & \textbf{Rec} & \textbf{F1} \\
\midrule
\multirow{3}{*}{ModernBERT} & \multirow{3}{*}{SH} & Irrelevant & .899 & .898 & .899 & .899 \\
 & & Partial & .870 & .816 & .870 & .842 \\
 & & Relevant & .904 & .900 & .904 & .902 \\
\midrule
\multirow{3}{*}{ModernBERT} & \multirow{3}{*}{MHP} & Irrelevant & .893 & .899 & .893 & .896 \\
 & & Partial & .857 & .848 & .857 & .852 \\
 & & Relevant & .900 & .903 & .900 & .902 \\
\midrule
\multirow{3}{*}{Llama-3-8b} & \multirow{3}{*}{MHP} & Irrelevant & .890 & .905 & .890 & .897 \\
 & & Partial & .855 & .840 & .855 & .847 \\
 & & Relevant & .903 & .904 & .903 & .904 \\
\midrule
\multirow{3}{*}{Gemma-2-2b} & \multirow{3}{*}{MHP} & Irrelevant & .893 & .903 & .893 & .898 \\
 & & Partial & .840 & .847 & .840 & .844 \\
 & & Relevant & .908 & .891 & .908 & .899 \\
\bottomrule
\end{tabular}
\end{table}

\section{Ablation Studies}
\label{sec:ablations}

We conduct ablation studies on four factors: (1)~dataset scale, numerical precision, and input format; (2)~class weights; (3)~model scale; and (4)~task routing architecture. The task routing ablation (Section~\ref{sec:routing_ablation}) is the most informative, as it surfaces the encoder--decoder asymmetry that is our central finding.

\subsection{Dataset Scale, Precision, and Input Format}

Doubling the training data yields +1.65\% to +3.33\% improvement, while float32 provides negligible gains over float16 (+0.02\% to +0.23\%). Table~\ref{tab:ablation_combined} summarizes both effects. The larger data gain for Gemma-2-2b suggests that smaller models benefit more from additional data. We recommend float16/bfloat16 for production deployments given the $\sim$2$\times$ training time cost of float32. We also compared task-prefix format (\texttt{``Task X Relevance: q1 [SEP] q2''}) against natural language question format (\texttt{``How relevant is `q2' to `q1'?''}) for Gemma-2-2b and found a negligible difference (87.16\% vs.\ 87.14\%); we adopt the task-prefix format for simplicity.

\begin{table}[t]
\centering
\small
\caption{Ablation: dataset scale and numerical precision on 453K test examples. Accuracy (\%). Differences $<$0.5\% are not significant.}
\label{tab:ablation_combined}
\begin{tabular}{@{}llccc@{}}
\toprule
\textbf{Factor} & \textbf{Model} & \textbf{Small (680K)} & \textbf{Large (1.36M)} & \textbf{$\Delta$} \\
\midrule
\multirow{2}{*}{Dataset scale} & Gemma-2-2b & 83.83 & 87.16 & +3.33 \\
 & Llama-3-8b & 86.22 & 87.87 & +1.65 \\
\midrule
 & & \textbf{fp16} & \textbf{fp32} & \\
\midrule
\multirow{2}{*}{Num.\ precision} & Gemma-2-2b & 87.14 & 87.16 & +0.02 \\
 & Llama-3-8b & 87.64 & 87.87 & +0.23 \\
\bottomrule
\end{tabular}
\end{table}

\begin{table}[t]
\centering
\small
\caption{Effect of class weights on Llama-3-8b (MHP, fp16) on 453K test examples. Per-class accuracy (\%). Differences $<$0.5\% are not significant.}
\label{tab:ablation_weights}
\begin{tabular}{@{}lcccc@{}}
\toprule
\textbf{Weights $\mathbf{w}$} & \textbf{Overall} & \textbf{Class 0} & \textbf{Class 1} & \textbf{Class 2} \\
\midrule
$[1.0, 1.0, 1.0]$ & 87.64 & 89.25 & 82.70 & 90.97 \\
$[1.0, 1.25, 1.0]$ & 87.12 & 86.75 & 84.36 & 90.24 \\
$[1.0, 1.5, 1.0]$ & 87.12 & 85.33 & \textbf{89.04} & 86.98 \\
\bottomrule
\end{tabular}
\end{table}

\subsection{Class Weights and Effect of Model Scale}
Upweighting the ``Partially Relevant'' class by $1.5\times$ improves its accuracy by +6.34\%, but costs $-3.99\%$ on ``Relevant'' (Table~\ref{tab:ablation_weights}). The moderate weighting $[1.0, 1.25, 1.0]$ provides a balanced trade-off and is our default for cross-encoder models.

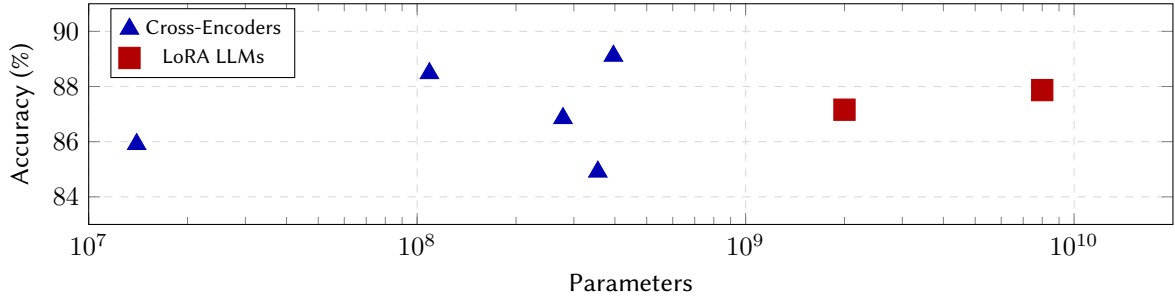
\begin{figure}[t]
\centering
\begin{tikzpicture}
\begin{semilogxaxis}[
    width=\columnwidth,
    height=4.5cm,
    xlabel={Parameters},
    ylabel={Accuracy (\%)},
    xmin=10000000, xmax=20000000000,
    ymin=83, ymax=91,
    grid=major,
    grid style={dashed, gray!30},
    legend style={at={(0.02,0.98)}, anchor=north west, font=\scriptsize},
    mark size=3pt,
]
\addplot[only marks, mark=triangle*, blue!70!black, mark size=4pt] coordinates {
    (14000000, 85.91)
    (109000000, 88.48)
    (278000000, 86.85)
    (355000000, 84.91)
    (396000000, 89.11)
};
\addplot[only marks, mark=square*, red!70!black, mark size=4pt] coordinates {
    (2000000000, 87.16)
    (8000000000, 87.87)
};
\legend{Cross-Encoders, LoRA LLMs}
\end{semilogxaxis}
\end{tikzpicture}
\caption{Model parameters (log scale) vs.\ accuracy. Cross-encoders show non-monotonic scaling; pre-training objective matters more than parameter count. LoRA LLMs train ${\sim}$0.1\% of total parameters.}
\label{fig:model_scale}
\end{figure}

Figure~\ref{fig:model_scale} shows that model scale does not monotonically improve accuracy for cross-encoders. RoBERTa-large (355M) achieves only 84.91\%, underperforming the much smaller ELECTRA-base (109M, 88.48\%). ModernBERT (396M) achieves the best cross-encoder accuracy (89.11\%), benefiting from its modern architecture with rotary embeddings and flash attention rather than raw parameter count. We hypothesize that RoBERTa's lower accuracy reflects its generic web-text pretraining, whereas ELECTRA's replaced-token-detection objective and ModernBERT's reranker initialization provide stronger inductive biases for fine-grained relevance discrimination. Among LoRA-adapted LLMs, the 2B-parameter Gemma achieves 87.16\%---within 0.71\% of the 8B Llama---suggesting that parameter efficiency can be achieved without substantial accuracy loss.

\subsection{Effect of Task Routing Architecture}
\label{sec:routing_ablation}

Removing text prefixes without private layers causes training collapse for decoder-only LLMs but barely affects cross-encoders---consistent with an encoder--decoder asymmetry in task identity encoding. Table~\ref{tab:routing_ablation} isolates this effect by comparing the same backbone model under three routing architectures.

\begin{table}[t]
\centering
\small
\caption{Task routing ablation. Overall accuracy (\%) and $\Delta$ vs.\ text-prefix baseline (SH). $^\dagger$Training collapse. All $|\Delta| \geq 0.5\%$ significant at $p < 0.01$.}
\label{tab:routing_ablation}
\begin{tabular}{@{}llcr@{}}
\toprule
\textbf{Model} & \textbf{Routing} & \textbf{Overall (\%)} & \textbf{$\Delta$ vs.\ SH} \\
\midrule
\multirow{3}{*}{ModernBERT} & SH (text-prefix) & 89.26 & -- \\
 & MH (multi-head) & 88.24 & $-1.02$ \\
 & MHP (MH+private) & 88.31 & $-0.95$ \\
\midrule
\multirow{3}{*}{Llama-3-8b} & SH (text-prefix) & 87.88 & -- \\
 & MH (multi-head) & 80.96 & $-6.92$ \\
 & MHP (MH+private) & 88.27 & $+0.39$ \\
\midrule
\multirow{3}{*}{Gemma-2-2b} & SH (text-prefix) & 87.14 & -- \\
 & MH (multi-head)$^\dagger$ & 55.59 & $-31.55$ \\
 & MHP (MH+private) & 88.04 & $+0.90$ \\
\bottomrule
\end{tabular}
\end{table}

\textbf{Cross-encoders are robust to routing changes.} ModernBERT MH (88.24\%) is within 1.02\% of its SH baseline (89.26\%), and MHP (88.31\%) narrows this gap to 0.95\%. The bidirectional encoder's [CLS] token naturally aggregates global context, so task-specific information encoded in text prefixes provides only marginal benefit over an explicit routing mechanism.

\textbf{Decoder-only LLMs are sensitive to routing architecture.} Gemma-2-2b MH collapses to 55.59\%---a 31.55\% drop from its SH baseline. Llama-3-8b MH degrades by 6.92\%. In both cases, replacing text-prefix routing with a simple linear head is insufficient. The MH confusion matrix for Gemma (not shown) shows that the model predominantly predicts ``Partially Relevant'' for all inputs, indicating a failure to learn task-discriminative representations without text prefixes. We did not tune the learning rate separately for MH; both MH and SH used the same lr ($1{\times}10^{-5}$). The collapse may therefore be partly attributable to training instability rather than purely architectural limitations. However, the fact that MHP recovers performance with the \emph{same} training configuration (same lr, same epochs, same data) suggests the private layers are the critical factor, not hyperparameter tuning.

\textbf{Private layers recover performance for LLMs.} Adding two private transformer layers per task (MHP) fully recovers performance: Gemma MHP (88.04\%) exceeds its SH baseline by +0.90\%, and Llama MHP (88.27\%) exceeds its baseline by +0.39\%. This suggests that the private layers provide sufficient task-specific representation capacity to compensate for the removed text prefix signal. The MHP architecture adds 3.3\% parameters for Gemma-2-2b and 5.0\% for Llama-3-8b. To disentangle routing from capacity, we note that MH (which adds $<$0.01\% parameters via the linear heads) recovers ModernBERT performance to within 1.02\%, suggesting that the improvement is primarily from the routing mechanism rather than extra parameters alone.

\textbf{Interpretation.} We hypothesize that this encoder--decoder asymmetry arises from differences in pooling and pretraining objectives. Cross-encoders produce a dedicated [CLS] representation that is architecturally trained to summarize the entire input, making it relatively task-agnostic. Decoder-only LLMs pool from the \emph{last token}, whose representation is shaped by the causal attention mask and heavily influenced by the sequence prefix. When the task-identifying prefix is removed, the last-token representation loses its primary source of task conditioning, and a linear projection alone cannot recover this information from the shared backbone.

\section{Analysis and Discussion}
\label{sec:discussion}

\noindent\textbf{Why does the metric choice reverse model rankings?}
ModernBERT SH leads on sample-weighted overall accuracy (89.26\%), but the MHP Ensemble achieves the highest result (89.96\%) and also leads on macro-averaged accuracy (86.19\%). The reason is that the test set is dominated by high-resource tasks---T1, T4, and T6 together constitute 86\% of examples---on which cross-encoders excel. When all six tasks are weighted equally, the MHP Ensemble's balanced gains across all tasks prove decisive. For practitioners, the choice of metric should reflect whether all tasks are equally important (macro) or whether production traffic distribution matters (sample-weighted).

\noindent\textbf{How much does data scarcity affect transfer magnitude?}
Following Standley et al.~\cite{standley2020tasks}, we hypothesized that multi-task training would yield positive transfer from data-rich to data-scarce tasks through shared semantic structure, but that negative transfer could arise when task objectives conflict---for instance, Task~6 (Ad--Ad) relies heavily on lexical overlap, while Task~3 (Query--PT) requires abstract categorical reasoning. The empirical results support this hypothesis. Multi-task training gains are inversely correlated with task training set size. For Gemma-2-2b (SH): (a)~T5 (14K train) gains +14.07\% (69.82\% $\to$ 83.89\%); (b)~T2 (143K train) gains +7.22\% (85.23\% $\to$ 92.45\%); (c)~T1 (404K train) loses $-4.75\%$ (86.71\% $\to$ 81.96\%); and (d)~T3 (34K train) shows the lowest accuracy across all configurations, reflecting inherent task difficulty rather than data scarcity alone. The negative transfer on T1 reflects the capacity trade-off inherent in multi-task training: performance on the largest task may decrease slightly as the model allocates capacity to smaller tasks. The MHP Ensemble mitigates this through model diversity, achieving the best T1 accuracy (84.64\%) among multi-task configurations.

\noindent\textbf{Why do encoder and decoder models respond differently to routing changes?}
Cross-encoders exhibit high per-task variance (ModernBERT SH range: 62.30\%--97.81\%), excelling on lexically-rich tasks through bidirectional attention but struggling on abstract semantic tasks (T3 Q--PT: 62.30\%). The routing ablation (Section~\ref{sec:routing_ablation}) is consistent with an encoder--decoder asymmetry: cross-encoders lose only $\sim$1\% when switching from text prefixes to multi-head routing, while decoder-only LLMs lose 7--32\%. This asymmetry has direct implications: (a)~for encoder-based models, text-prefix routing is a simple and effective default, and multi-head routing provides a marginal trade-off of 1\% accuracy for cleaner architectural separation; (b)~for decoder-only LLMs, private layers are architecturally necessary when removing text prefixes, as the 2-layer private stack provides sufficient task-specific capacity to match or exceed text-prefix performance. The MHP Ensemble (89.96\%) outperforms the SH Ensemble (88.04\%) by 1.92\%, suggesting that private-layer routing induces more diverse per-model error patterns that the ensemble can exploit.

\noindent\textbf{What are the practical implications for deployment?}
For production deployment, the MHP architecture offers cleaner separation of shared and task-specific parameters, which could facilitate independent task updates without retraining the shared backbone; Gemma-2-2b MHP (88.04\%) achieves the best accuracy-efficiency trade-off for latency-sensitive applications. At inference time, ModernBERT-396M processes ${\sim}$2,500 examples/sec on a V100 GPU (batch size 64, sequence length 128), while Gemma-2-2b with LoRA processes ${\sim}$180 examples/sec (batch size 16, sequence length 64). The MHP private layers add negligible overhead ($<$2\% latency increase) since they consist of only 2 transformer layers atop the shared backbone. For offline annotation, ModernBERT SH (89.26\%) remains preferred for batch annotation of lexically-rich tasks (T6 Ad--Ad: 97.81\%), while the MHP Ensemble (89.96\%) is optimal when all tasks matter equally. The ensemble requires running three models at inference time, tripling serving cost for a +0.70\% gain over the best single model; whether this trade-off is worthwhile depends on the application's accuracy requirements. For ensembling, the MHP configuration achieves sufficient diversity (16.01\% disagreement rate) to correct individual model errors while maintaining high agreement on easy examples.

\noindent\textbf{What do the remaining errors look like?}
Qualitative inspection of 100 randomly sampled errors from the MHP Ensemble reveals three dominant failure modes: (a)~partial relevance misclassified as irrelevant when entity pairs share taxonomy but differ in specificity (e.g., ``running shoes'' vs.\ Athletic Footwear), accounting for ${\sim}$38\% of sampled errors; (b)~irrelevant pairs misclassified as partially relevant when product titles contain query terms in non-relevant contexts (e.g., ``apple'' the fruit vs.\ Apple brand accessories), accounting for ${\sim}$29\%; and (c)~cross-task confusion on T3 (Query--PT) where abstract categorical mappings have no lexical overlap, accounting for ${\sim}$22\%. The remaining ${\sim}$11\% are borderline cases where even human annotators disagree. These proportions are approximate given the 100-example sample.

\noindent\textbf{Limitations.}
We note the following limitations: (a)~our dataset is proprietary and cannot be publicly released, though the methodology is applicable to any e-commerce platform with human-labeled relevance data; (b)~LoRA and cross-encoder models use different hyperparameters (sequence length, epochs, class weights), confounding the pure architectural comparison---the routing ablation controls for this by comparing the same backbone under different routing strategies; (c)~the Gemma MH training collapse may be addressable with warmup, gradient clipping, or curriculum learning, but we report MH as-is to document the failure mode, since MHP provides a clean solution; (d)~despite human validation showing $>$82\% agreement (Table~\ref{tab:validation}), residual label noise remains in Tasks 2--6; (e)~our evaluation is English-only with a three-point scale; the three-point scale reflects production annotation guidelines that prioritize annotator agreement ($\kappa \geq 0.76$) over granularity, and extension to a four-point scale (e.g., ESCI~\cite{reddy2022shopping}) and multilingual settings is left for future work; and (f)~we do not evaluate on the public ESCI benchmark because our three-point scale and six-task formulation differ from ESCI's four-point single-task setup; direct comparison would require label re-mapping, which we leave for future work.

\section{Conclusion}
\label{sec:conclusion}

How should practitioners build a unified relevance model that spans multiple entity-pair tasks in e-commerce? In this paper, we addressed this question by developing a multi-task relevance framework and conducting the first systematic comparison of task routing architectures across encoder-based and decoder-only model families.

Three findings stand out. (a)~The choice of task routing architecture is not neutral: private-layer routing (MHP) enables the strongest ensemble (89.96\%, +1.92\% over text-prefix), while the simpler multi-head approach (MH) collapses for decoder-only LLMs but not for cross-encoders. (b)~This encoder--decoder asymmetry---cross-encoders lose only 1\% when text prefixes are replaced by multi-head routing, while Gemma-2-2b loses 31.6\%---has direct practical implications for architecture selection. (c)~Multi-task training lifts low-resource tasks by up to +14\%, and the compact 2B-parameter Gemma MHP (88.04\%) achieves within 0.23\% of the 8B Llama MHP, suggesting that parameter efficiency need not sacrifice accuracy.

Overall, our work provides a practical blueprint for building unified multi-task relevance models in e-commerce and suggests that the choice of task routing architecture is not neutral---it interacts with the model family in ways that can cause training collapse or enable stronger ensembles. Having shown that hard deterministic routing suffices for our six-task setting (Section~\ref{sec:routing}), future work will investigate (1)~whether learned soft routing (e.g., MoE gating) provides further gains, (2)~knowledge distillation from the MHP Ensemble into a single student model for real-time serving, (3)~online A/B testing of the multi-head architecture in production, and (4)~extension to multilingual settings and finer relevance granularity.

\section*{Declaration on Generative AI}
During the preparation of this work, the author(s) used Generative AI only for formatting assistance in plotting code used to produce figures, and not for drafting, analysis, interpretation, or any other part of the manuscript. The author(s) reviewed, corrected, and validated the generated code and figures and take(s) full responsibility for the publication's content.

\bibliography{references}

@inproceedings{rokon2024enhancement,
  title={Enhancement of E-commerce Sponsored Search Relevancy with {LLM}},
  author={Rokon, Md Omar Faruk and Simion, Andrei and Du, Weizhi and Wen, Musen and Yao, Hong and Lee, Kuang-chih},
  booktitle={Proceedings of the SIGIR Workshop on eCommerce (eCom'24)},
  year={2024}
}

@article{desai2026unified,
      title={Unified Supervision for Walmarts Sponsored Search Retrieval via Joint Semantic Relevance and Behavioral Engagement Modeling}, 
      author={Desai, Shasvat and Rokon, Md Omar Faruk  and Acharya, Jhalak Nilesh and Shah, Isha and Yao, Hong and Porwal, Utkarsh and Lee, Kuang-chih},
      year={2026},
      journal={arXiv preprint arXiv:2604.07930}
}

@inproceedings{hu2021lora,
  title={{LoRA}: Low-Rank Adaptation of Large Language Models},
  author={Hu, Edward J and Shen, Yelong and Wallis, Phillip and Allen-Zhu, Zeyuan and Li, Yuanzhi and Wang, Shean and Wang, Lu and Chen, Weizhu},
  booktitle={Proceedings of the Tenth International Conference on Learning Representations (ICLR)},
  year={2022}
}

@article{liu2019mtdnn,
  title={Multi-Task Deep Neural Networks for Natural Language Understanding},
  author={Liu, Xiaodong and He, Pengcheng and Chen, Weizhu and Gao, Jianfeng},
  journal={arXiv preprint arXiv:1901.11504},
  year={2019}
}

@inproceedings{standley2020tasks,
  title={Which Tasks Should Be Learned Together in Multi-task Learning?},
  author={Standley, Trevor and Zamir, Amir R and Chen, Dawn and Guibas, Leonidas and Malik, Jitendra and Savarese, Silvio},
  booktitle={International Conference on Machine Learning},
  pages={9120--9132},
  year={2020}
}

@article{caruana1997multitask,
  title={Multitask Learning},
  author={Caruana, Rich},
  journal={Machine Learning},
  volume={28},
  number={1},
  pages={41--75},
  year={1997}
}

@inproceedings{reddy2022shopping,
  title={Shopping Queries Dataset: A Large-Scale {ESCI} Benchmark for Improving Product Search},
  author={Reddy, Chandan K and M{\`a}rquez, Llu{\'\i}s and Valero, Fran and Rao, Nikhil and Zaragoza, Hugo and Bandyopadhyay, Sambaran and Biswas, Arnab and Xing, Anlu and Subbian, Karthik},
  booktitle={Proceedings of the 28th ACM SIGKDD Conference on Knowledge Discovery and Data Mining},
  pages={4429--4439},
  year={2022}
}

@inproceedings{aiello2016relevance,
  title={The Role of Relevance in Sponsored Search},
  author={Aiello, Luca Maria and Arapakis, Ioannis and Baeza-Yates, Ricardo and Bai, Xiao and Barbieri, Nicola and Mantrach, Amin and Silvestri, Fabrizio},
  booktitle={Proceedings of the 25th ACM International Conference on Information and Knowledge Management},
  pages={185--194},
  year={2016}
}

@article{jarvelin2002cumulated,
  title={Cumulated Gain-Based Evaluation of {IR} Techniques},
  author={J{\"a}rvelin, Kalervo and Kek{\"a}l{\"a}inen, Jaana},
  journal={ACM Transactions on Information Systems},
  volume={20},
  number={4},
  pages={422--446},
  year={2002}
}

@article{nogueira2019passage,
  title={Passage Re-ranking with {BERT}},
  author={Nogueira, Rodrigo and Cho, Kyunghyun},
  journal={arXiv preprint arXiv:1901.04085},
  year={2019}
}

@article{devlin2019bert,
  title={{BERT}: Pre-training of Deep Bidirectional Transformers for Language Understanding},
  author={Devlin, Jacob and Chang, Ming-Wei and Lee, Kenton and Toutanova, Kristina},
  journal={arXiv preprint arXiv:1810.04805},
  year={2019}
}

@article{team2024gemma,
  title={Gemma: Open Models Based on {Gemini} Research and Technology},
  author={{Gemma Team}},
  journal={arXiv preprint arXiv:2403.08295},
  year={2024}
}

@article{grattafiori2024llama3,
  title={The {Llama 3} Herd of Models},
  author={Grattafiori, Aaron and Dubey, Abhimanyu and Jauhri, Abhinav and others},
  journal={arXiv preprint arXiv:2407.21783},
  year={2024}
}

@article{achiam2023gpt4,
  title={{GPT-4} Technical Report},
  author={Achiam, Josh and Adler, Steven and Agarwal, Sandhini and Ahmad, Lama and Akkaya, Ilge and Aleman, Florencia Leoni and Almeida, Diogo and others},
  journal={arXiv preprint arXiv:2303.08774},
  year={2023}
}

@article{warner2024modernbert,
  title={Smarter, Better, Faster, Longer: A Modern Bidirectional Encoder for Fast, Memory Efficient, and Long Context Finetuning and Inference},
  author={Warner, Benjamin and Chaffin, Antoine and Clavi{\'e}, Benjamin and Weller, Orion and Hallstr{\"o}m, Oskar and Taghadouini, Said and Gallagher, Alexis and Biswas, Raja and Ladhak, Faisal and Aarsen, Tom and Cooper, Nathan and Adams, Griffin and Howard, Jeremy and Poli, Iacopo},
  journal={arXiv preprint arXiv:2412.13663},
  year={2024}
}

@article{reimers2019sentencebert,
  title={Sentence-{BERT}: Sentence Embeddings using Siamese {BERT}-Networks},
  author={Reimers, Nils and Gurevych, Iryna},
  journal={arXiv preprint arXiv:1908.10084},
  year={2019}
}

@inproceedings{clark2020electra,
  title={{ELECTRA}: Pre-training Text Encoders as Discriminators Rather Than Generators},
  author={Clark, Kevin and Luong, Minh-Thang and Le, Quoc V and Manning, Christopher D},
  booktitle={International Conference on Learning Representations},
  year={2020}
}

@article{jiao2020tinybert,
  title={{TinyBERT}: Distilling {BERT} for Natural Language Understanding},
  author={Jiao, Xiaoqi and Yin, Yichun and Shang, Lifeng and Jiang, Xin and Chen, Xiao and Li, Linlin and Wang, Fang and Liu, Qun},
  journal={arXiv preprint arXiv:1909.10351},
  year={2020}
}

@article{liu2019roberta,
  title={{RoBERTa}: A Robustly Optimized {BERT} Pretraining Approach},
  author={Liu, Yinhan and Ott, Myle and Goyal, Naman and Du, Jingfei and Joshi, Mandar and Chen, Danqi and Levy, Omer and Lewis, Mike and Zettlemoyer, Luke and Stoyanov, Veselin},
  journal={arXiv preprint arXiv:1907.11692},
  year={2019}
}

@article{xiao2024bge,
  title={{BGE} {M3}-Embedding: Multi-Lingual, Multi-Functionality, Multi-Granularity Text Embeddings Through Self-Knowledge Distillation},
  author={Chen, Jianlv and Xiao, Shitao and Zhang, Peitian and Luo, Kun and Lian, Defu and Liu, Zheng},
  note={\url{https://github.com/FlagOpen/FlagEmbedding}},
  journal={arXiv preprint arXiv:2402.03216},
  year={2024}
}

@inproceedings{thomas2024large,
  title={Large Language Models Can Accurately Predict Searcher Preferences},
  author={Thomas, Paul and Spielman, Seth and Craswell, Nick and Mitra, Bhaskar},
  booktitle={Proceedings of the 47th International ACM SIGIR Conference on Research and Development in Information Retrieval},
  pages={1930--1940},
  year={2024}
}

@inproceedings{faggioli2023perspectives,
  title={Perspectives on Large Language Models for Relevance Judgment},
  author={Faggioli, Guglielmo and Dietz, Laura and Clarke, Charles and Demartini, Gianluca and Hagen, Matthias and Hauff, Claudia and Kando, Noriko and Kanoulas, Evangelos and Potthast, Martin and Stein, Benno and Wachsmuth, Henning},
  booktitle={Proceedings of the 2023 ACM SIGIR International Conference on the Theory of Information Retrieval},
  pages={39--50},
  year={2023}
}

@article{dettmers2023qlora,
  title={{QLoRA}: Efficient Finetuning of Quantized {LLMs}},
  author={Dettmers, Tim and Pagnoni, Artidoro and Holtzman, Ari and Zettlemoyer, Luke},
  journal={arXiv preprint arXiv:2305.14314},
  year={2023}
}

@inproceedings{cormack2009reciprocal,
  title={Reciprocal Rank Fusion Outperforms {Condorcet} and Individual Rank Learning Methods},
  author={Cormack, Gordon V and Clarke, Charles LA and Buettcher, Stefan},
  booktitle={Proceedings of the 32nd International ACM SIGIR Conference on Research and Development in Information Retrieval},
  pages={758--759},
  year={2009}
}

@inproceedings{wang2023clickconv,
  title={Click-Conversion Multi-Task Model with Position Bias Mitigation for Sponsored Search in Ecommerce},
  author={Wang, Yibo and Xue, Yanbing and Liu, Bo and Wen, Musen and Zhao, Wenting and Guo, Stephen and Yu, Philip S},
  booktitle={Proceedings of the 46th International ACM SIGIR Conference on Research and Development in Information Retrieval},
  year={2023}
}

@inproceedings{su2018user,
  title={User Intent, Behaviour, and Perceived Satisfaction in Product Search},
  author={Su, Ning and He, Jiyin and Liu, Yiqun and Zhang, Min and Ma, Shaoping},
  booktitle={Proceedings of the Eleventh ACM International Conference on Web Search and Data Mining},
  pages={547--555},
  year={2018}
}

@article{loshchilov2019adamw,
  title={Decoupled Weight Decay Regularization},
  author={Loshchilov, Ilya and Hutter, Frank},
  journal={arXiv preprint arXiv:1711.05101},
  year={2019}
}

@article{ruder2017overview,
  title={An Overview of Multi-Task Learning in Deep Neural Networks},
  author={Ruder, Sebastian},
  journal={arXiv preprint arXiv:1706.05098},
  year={2017}
}

@article{crawshaw2020multitask,
  title={Multi-Task Learning with Deep Neural Networks: A Survey},
  author={Crawshaw, Michael},
  journal={arXiv preprint arXiv:2009.09796},
  year={2020}
}

@inproceedings{ma2018mmoe,
  title={Modeling Task Relationships in Multi-task Learning with Multi-gate Mixture-of-Experts},
  author={Ma, Jiaqi and Zhao, Zhe and Yi, Xinyang and Chen, Jilin and Hong, Lichan and Chi, Ed H},
  booktitle={Proceedings of the 24th ACM SIGKDD International Conference on Knowledge Discovery and Data Mining},
  pages={1930--1939},
  year={2018}
}

@article{cohen1960kappa,
  title={A Coefficient of Agreement for Nominal Scales},
  author={Cohen, Jacob},
  journal={Educational and Psychological Measurement},
  volume={20},
  number={1},
  pages={37--46},
  year={1960}
}

@inproceedings{dietterich2000ensemble,
  title={Ensemble Methods in Machine Learning},
  author={Dietterich, Thomas G},
  booktitle={International Workshop on Multiple Classifier Systems},
  pages={1--15},
  year={2000}
}

@article{wang2024semantic,
  title={Semantic Ads Retrieval at {Walmart} eCommerce with Language Models Progressively Trained on Multiple Knowledge Domains},
  author={Wang, Zhaodong and Du, Weizhi and Rokon, Md Omar Faruk and Adhikary, Pooshpendu and Xue, Yanbing and Xu, Jiaxuan and Zhou, Jianghong and Lee, Kuang-chih and Wen, Musen},
  journal={arXiv preprint arXiv:2502.09089},
  year={2025}
}

@inproceedings{rao2020product,
  title={Product Insights: Analyzing Product Intents in Web Search},
  author={Rao, Nikitha and Bansal, Chetan and Mukherjee, Subhabrata and Maddila, Chandra},
  booktitle={Proceedings of the 29th ACM International Conference on Information and Knowledge Management},
  pages={2189--2192},
  year={2020}
}

@inproceedings{tang2020ple,
  title={Progressive Layered Extraction ({PLE}): A Novel Multi-Task Learning ({MTL}) Model for Personalized Recommendations},
  author={Tang, Hongyan and Liu, Junning and Zhao, Ming and Gong, Xudong},
  booktitle={Proceedings of the 14th ACM Conference on Recommender Systems},
  pages={269--278},
  year={2020}
}

@String{Computing = "Computing" }

@String{Computer = "{IEEE} Computer" }

@BOOK{test,
   author = "Donald E. Knuth",
   title = "Seminumerical Algorithms",
   volume = 2,
   series = "The Art of Computer Programming",
   publisher = "Addison-Wesley",
   address = "Reading, MA",
   edition = "2nd",
   month = "10~" # jan,
   year = "1981",
}

@misc{R,
    title = {R: A Language and Environment for Statistical Computing},
    author = {{R Core Team}},
    organization = {R Foundation for Statistical Computing},
    address = {Vienna, Austria},
    year = {2019},
    url = {https://www.R-project.org/},
}

\end{document}